\documentclass{appolb}
\usepackage{epsfig}
\usepackage{cite}

\begin{document}
\pagestyle{plain}
\title{ \bf TRANSVERSE SPIN EFFECTS IN $H/A\to \tau^+ \tau^-;~ 
\tau^\pm \to \nu X^\pm$, \\ MONTE CARLO APPROACH
\thanks{This work is partly supported by
the Polish State Committee for Scientific Research 
(KBN) grants Nos 
5P03B12420, 5P03B09320, 5P03B10121, 
and 
and the European Community's Human Potential
Programme under contract HPRN-CT-2000-00149 Physics at Colliders.} 
~\thanks{{\tt hep-ph/0202007}, {\tt LC-PHSM-2003-049}}
}
\author{  Zbigniew W\c{a}s 
\address{Institute of Nuclear Physics\\
         Kawiory 26a, 30-055 Cracow, Poland \\CERN, Theory Division, CH-1211 Geneva 23, Switzerland }
\and
Ma\l gorzata Worek
\address{Institute of  Physics, University of Silesia \\ Uniwersytecka 4,
 40-007 Katowice, Poland }
}
\maketitle
\vspace{-0.7 cm}
\begin{abstract}
The  transverse spin effects may be helpful to distinguish between 
scalar $(J^{PC}=0^{++})$ or pseudoscalar $(J^{PC}=0^{-+})$
nature of the spin zero (Higgs) particle once discovered in future accelerator
experiments. 
The correlations can manifest themselves
{\it e.g.} in the distribution of acollinearity angle of  $ X^\pm $ in 
the decay chain
$H/A\to \tau^+ \tau^-;~ \tau^\pm \to \nu X^\pm $. This delicate measurement 
will require however reconstruction of the Higgs boson rest-frame. Then,
questions of the combined detection-theoretical effects
may be critical to establish the reliability of the method. 
An appropriate Monte Carlo program
is essential.\\
To enable such studies we have extended the standard universal interface, 
of the
{\tt TAUOLA} $\tau$-lepton decay library, to include the complete spin effects
for
$\tau$ leptons originating  from the spin zero particle.  
The  interface is expected to work with any
Monte Carlo generator providing Higgs boson production, and subsequent decay 
into a pair of $\tau$ leptons.\\
Examples of numerical results and cross checks of the program 
will be also given. In particular, we find that effects of beamstrahlung may 
be critical to the quality of the measurement of the Higgs boson, unless some 
improvements of the method can be found.

\end{abstract}

One of the main goals for future high energy experiments is to measure properties
of the Standard Model (${\cal SM}$) Higgs sector.
Proton-(Anti)Proton Colliders, such as Tevatron \cite{Tevatron}  or LHC \cite{LHC1,LHC2}
are expected to discover the Higgs boson,
if the ${\cal SM}$ or one of its ${\cal MSSM}$ extensions  is true. Otherwise the spectrum of
possibilities is practically unlimited and discovery can not be guaranteed. 
The comprehensive precise measurements  of all Higgs boson 
 properties are  expected to be  left for future experiments
on high energy $e^+e^-$ linear colliders such as   JLC \cite{Abe:2001gc},
NLC \cite{:2001ve}, or TESLA \cite{TESLA}. 

One of the important measurement, just after establishing that 
the newly discovered 
particle has indeed spin zero, is to check  if it is a scalar or pseudoscalar. 
Depending on the mass of the (to be) discovered Higgs boson, and 
if it is  of Standard Model  or one of its numerous extensions,
different observables can give access to this information.
Already a long
time ago, see  {\it e.g.} \cite{Kramer:1994jn}, it was argued that
exploring transverse spin correlations in 
the Higgs boson decay  $H/A \to \tau^+ \tau^-; \tau^\pm \to \nu_{\tau} X^\pm$ can 
in some cases provide a {\it model independent} ~test. The method relies on the properties of the
Higgs boson Yukawa coupling to the $\tau$ lepton, which in the general case can be written  as 
$ \bar \tau (a_\tau+ i b_\tau \gamma_5) \tau$ (for a discussion of the Higgs boson models, see {\it e.g.} \cite{Abe:2001nn} page 123). The method,
 at least in  principle does not depend on the Higgs boson 
production mechanism at all. 

 There are many
reasons why this process  may turn out not to be interesting. The cross section may be 
too small for the luminosity of the future collider, the mass of the Higgs boson may be 
  heavy and
other Higgs boson decay channels  better suited for
the parity measurement. Finally the parity of the Higgs boson may be measurable  
from the properties
of its production.
However, it is generally accepted that 
the  $H/A \to \tau^+ \tau^-$ offers a very interesting signature.
Its feasibility needs to be studied especially in the context of decisions to be taken
on properties of the future LC detectors which  may be taken soon. The proposed 
 measurement  \cite{Kramer:1994jn} is {\it experimentally} involved. It requires 
reconstruction of the
acollinearity angle ($\delta^*$) between  $\tau^+$  $\tau^-$ decay 
products {\it in the H/A rest-frame}. Note that in  case of $H/A \to \tau^+ \tau^-$ 
the four momentum of the Higgs boson
is not directly measurable as we have the unobservable $\nu_\tau$ among its decay products; 
it needs to be reconstructed from
the constraints of energy momentum conservation for the whole event.
The distribution in  angle ($\delta^*$) is sensitive to the transverse $\tau^+ \tau^-$ spin correlations, 
which are different for the scalar, pseudoscalar or the mixed state 
(we will take into considerations only the extreme cases corresponding to choosing either 
$b_\tau$ or $a_\tau$ equal zero). 
Precise enough reconstruction of the {\it  H/A rest-frame} is important.
Many effects, theoretical ({\it e.g.} QED bremsstrahlung), or experimental (uncertainty
on beam energies, 
not sufficient
hermeticity of the detector, angular/energy resolution for all particles and jets {\it etc.}) 
may invalidate the method.
In the following, we will concentrate on the
feasibility of the method, taking into account properties of the
 $H/A \to \tau^+ \tau^-$ decay, and simple assumptions on bremsstrahlung and beamstrahlung 
in reconstructing Higgs boson rest-frame,
leaving out from the considerations all other limitations and constraints, be it theoretical 
or experimental.

 It is generally expected  
that the Monte Carlo method is the only way to estimate whether the measurement can be
realized in practice, and which features  
of the future detection set up may turn out to be crucial.  
Our paper is organized as follows.
First, an algorithm for generating decays of $\tau^\pm$ leptons produced
in $H/A \to \tau^+ \tau^-$ including full spin correlations for the Higgs boson production mechanism
is explained, and some numerical examples testing  the correctness of the program are given. 
Later, results taking into account inaccuracies in reconstructing the Higgs boson
rest-frame are shown and conclusions are given.

Since the Higgs boson spin is zero,
the spin correlations of its decay products {\it do not} depend at all 
on the mechanism of the Higgs boson production. 
Technical difficulties related to the choice of
$\tau^+$ and $\tau^-$ spin quantization frames, present in the case  
of  $e^+e^- \to  Z/\gamma \to \tau^+  \tau^-$ \cite{gps:1998,jadach-was:1984}
(bremsstrahlung effects included or not), are not present. The analytical form of the density
matrix is simple.
To calculate the density matrix for the pair of $\tau$-leptons  it is thus enough to: know 
their four momenta, know that they indeed originate from the Higgs boson and, 
assume the type of 
the Yukawa interaction. Such information is stored  in the event data structure
called  {\tt HEPEVT} common block \cite{PDG:1998} used by practically
 all Monte Carlo generators
for Higgs boson production. 

\begin{figure}[!ht]
\setlength{\unitlength}{0.1mm}
\begin{picture}(1600,800)
\put( 375,750){\makebox(0,0)[b]{\large }}
\put(1225,750){\makebox(0,0)[b]{\large }}
\put(30, -350){\makebox(0,0)[lb]{\epsfig{file=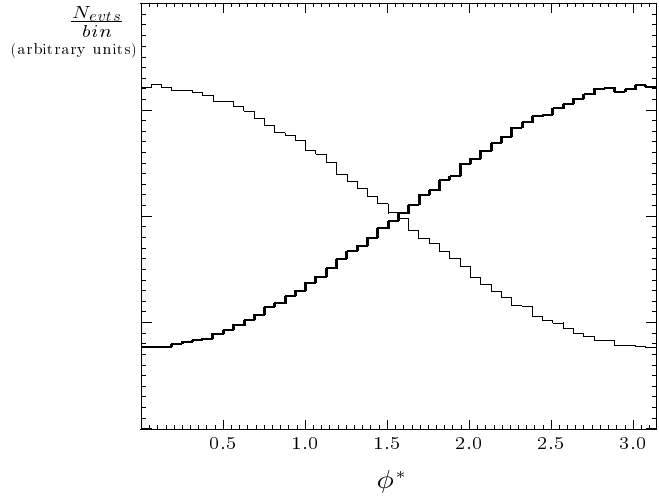,width=120mm,
height=140mm}}}
\end{picture}
\caption
{\it  The $\pi^+ \pi^-$ acoplanarity  distribution (angle $\phi^*$)  
in the  Higgs boson rest frame.   The thick line denotes the case 
of the scalar Higgs boson and 
thin line the pseudoscalar one.}
\label{rysunek1}
\end{figure}
\begin{figure}[!ht]
\setlength{\unitlength}{0.1mm}
\begin{picture}(1600,800)
\put( 375,750){\makebox(0,0)[b]{\large }}
\put(1225,750){\makebox(0,0)[b]{\large }}
\put(30, -350){\makebox(0,0)[lb]{\epsfig{file=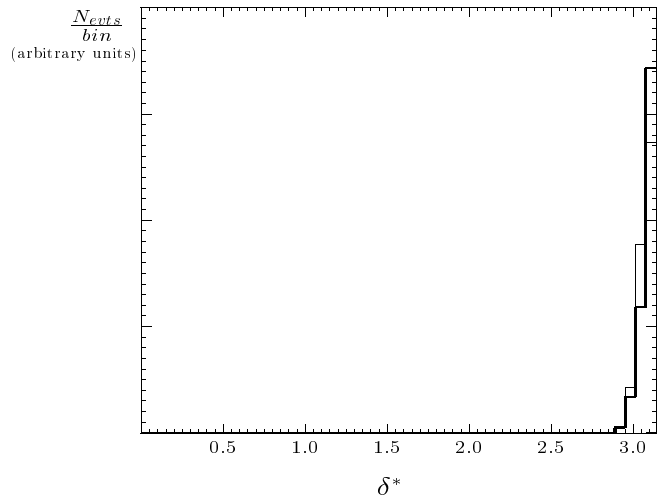,width=120mm,
height=140mm}}}
\end{picture}
\caption
{\it The  $\pi^+ \pi^-$ acollinearity distribution  (angle $\delta^*$)  
in the  Higgs boson rest frame. Full angular range $0 <\delta^* < \pi $ is shown.
The thick line denotes the case 
of the scalar Higgs boson and 
thin line the pseudoscalar one. }
\label{rysunek2}
\end{figure}
\begin{figure}[!ht]
\setlength{\unitlength}{0.1mm}
\begin{picture}(1600,800)
\put( 375,750){\makebox(0,0)[b]{\large }}
\put(1225,750){\makebox(0,0)[b]{\large }}
\put(-370,-1450){\makebox(0,0)[lb]{\epsfig{file=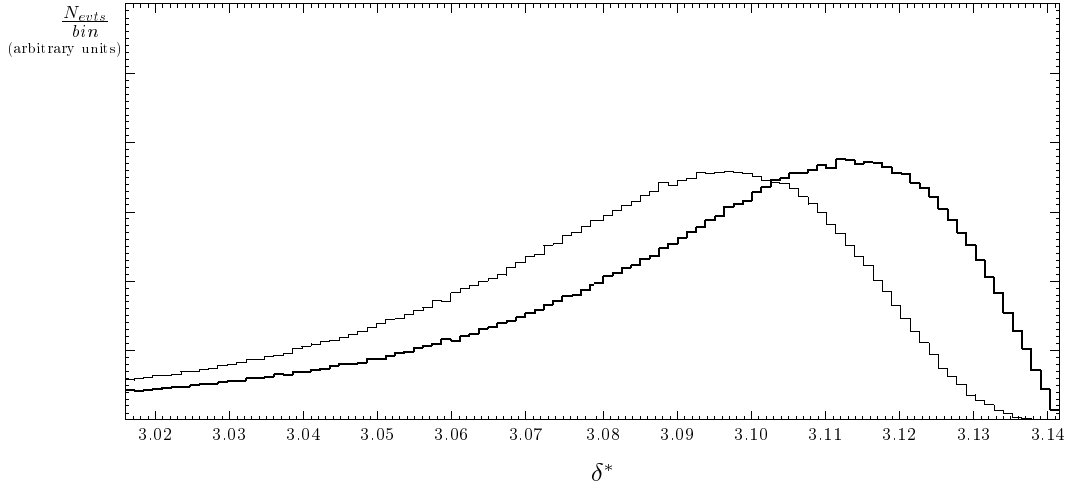,width=190mm,height=240mm}}}
\end{picture}
\caption
{\it The  $\pi^+ \pi^-$ acollinearity distribution  (angle $\delta^*$)  
in the  Higgs boson rest frame.
Parts of the distribution close to the end of the spectrum; $\delta^* \sim \pi$ are shown.
The thick line denotes the case 
of the scalar Higgs boson and the
thin line the pseudoscalar one. }
\label{rysunek3}
\end{figure}

\begin{figure}[!ht]
\setlength{\unitlength}{0.1mm}
\begin{picture}(1600,800)
\put( 375,750){\makebox(0,0)[b]{\large }}
\put(1225,750){\makebox(0,0)[b]{\large }}
\put(-30,-350){\makebox(0,0)[lb]{\epsfig{file=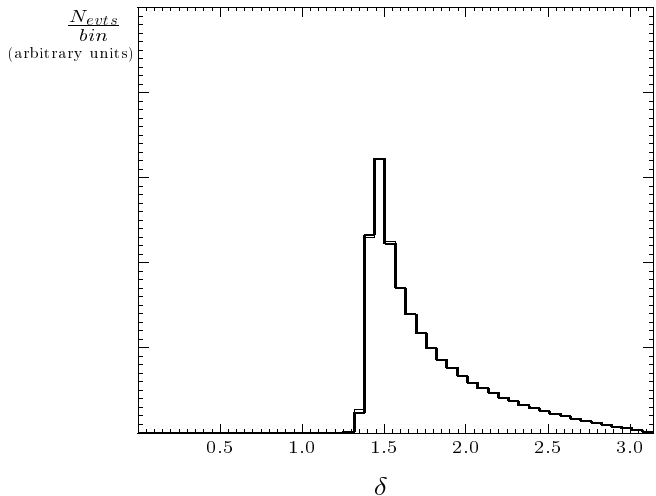,width=120mm,height=140mm}}}
\end{picture}
\caption
{\it  The  $\pi^+ \pi^-$ acollinearity distribution  (angle $\delta$)  
in the laboratory frame. Full angular range $0 <\delta< \pi $ is shown.
The thick line denotes the case when all spin effects are included in the decay 
of the scalar Higgs boson, while  only longitudinal spin correlations are included for 
thin line. The two lines are nearly 
indistinguishable.}
\label{rysunek5}
\end{figure}

\begin{figure}[!ht]
\setlength{\unitlength}{0.1mm}
\begin{picture}(1600,800)
\put( 375,750){\makebox(0,0)[b]{\large }}
\put(1225,750){\makebox(0,0)[b]{\large }}
\put(-30, -350){\makebox(0,0)[lb]{\epsfig{file=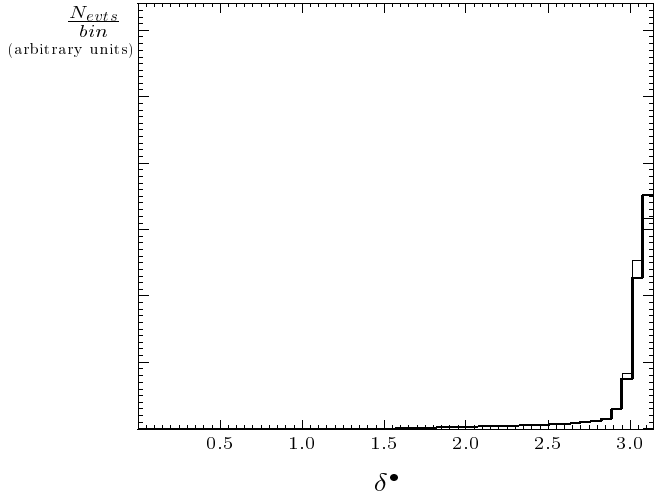,width=120mm,
height=140mm}}}
\end{picture}
\caption
{\it The  $\pi^+ \pi^-$ acollinearity distribution  (angle $\delta^\bullet$)  
in the scalar Higgs boson reconstructed rest frame. 
Full angular range $0 <\delta^\bullet < \pi $ is shown.
The thick line denotes the case when all spin effects are included, while 
only longitudinal spin correlations are taken for 
thin line.
 }
\label{rysunek6}
\end{figure}

\begin{figure}[!ht]
\setlength{\unitlength}{0.1mm}
\begin{picture}(1600,800)
\put( 375,750){\makebox(0,0)[b]{\large }}
\put(1225,750){\makebox(0,0)[b]{\large }}
\put(-350,-1550){\makebox(0,0)[lb]{\epsfig{file=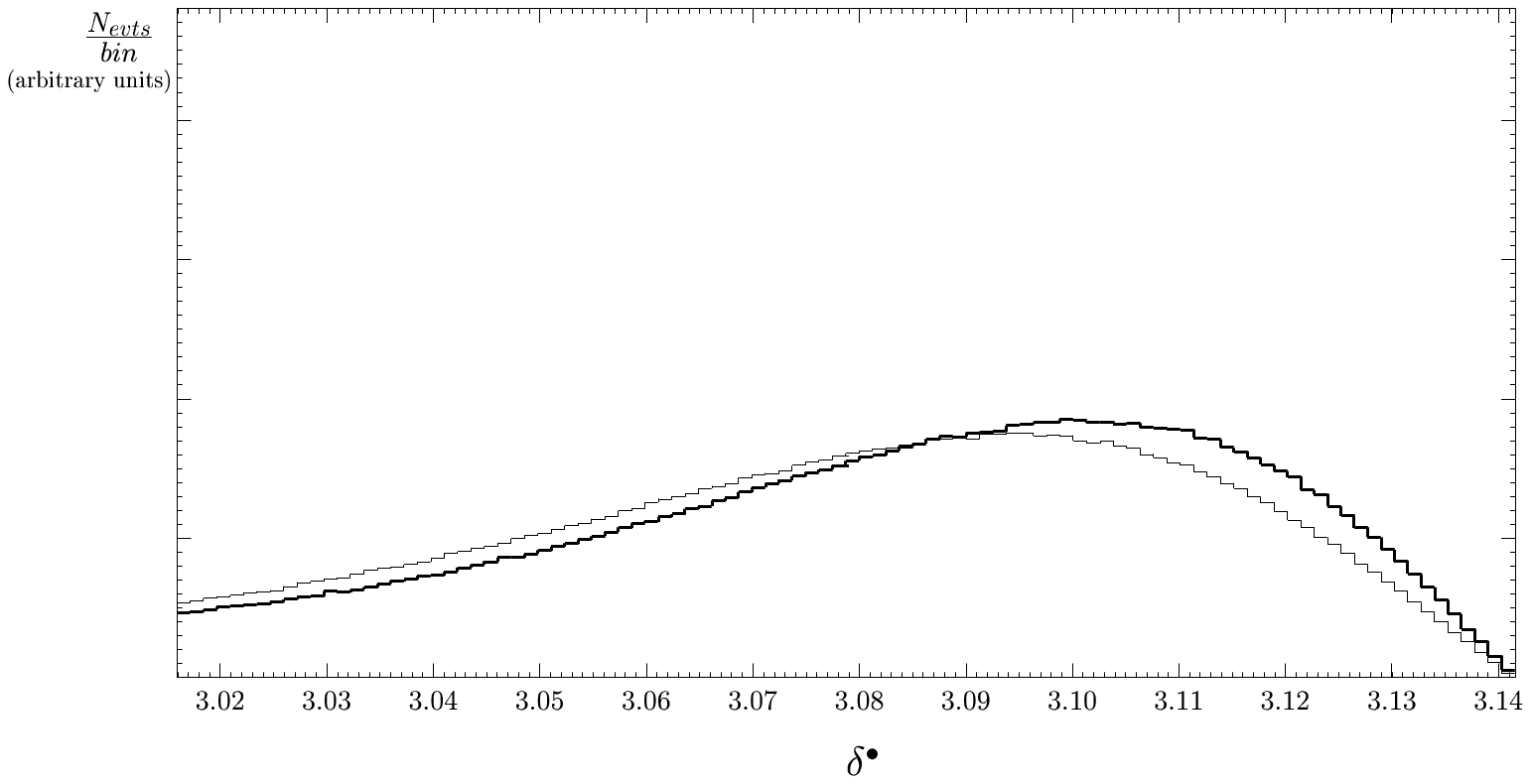,width=190mm,height=240mm}}}
\end{picture}
\caption
{\it  The  $\pi^+ \pi^-$ acollinearity distribution  (angle $\delta^\bullet$)  
in the scalar Higgs boson reconstructed rest frame. 
Parts of the distribution close to the end of the spectrum; $\delta^\bullet \sim \pi$ are shown.
The thick line denotes the case when all spin effects are included, while 
only longitudinal spin correlations are taken for 
thin line.}
\label{rysunek7}
\end{figure}

In  Refs. \cite{Pierzchala:2001gc,Golonka:2002iu}, the algorithm was developed where all
$ \tau$ leptons found in the {\tt HEPEVT} common block can be decayed 
with the help of the {\tt TAUOLA} library \cite{Jadach:1990mz,Jezabek:1991qp,Jadach:1993hs}
and the $ \tau$ decay products are appended to the {\tt HEPEVT} as well.
The kinematical information
on the momenta of all particles forming an event was used to calculate, in some approximation,
the longitudinal spin state of the $ \tau$.
For our purpose
that solution had to be extended, to incorporate the full density matrix of the $\tau^+ \tau^-$
pair, in the case  when it is originating  from the Higgs boson decay. The following changes
had to be introduced to the algorithm explained in Ref.~\cite{Pierzchala:2001gc}:

\begin{enumerate}
\item
The quantization frames for the spin states of 
$ \tau^+$ and  $\tau^-$ need to be  properly oriented with respect to each other. 
In our solution  they are simply connected by the boost along $\tau$ lepton momenta
as defined in the Higgs boson rest-frame. At the technical level this is enforced
by the  {\tt TRALOR} routine \cite{Jadach:1990mz}  defining the relation 
of the $\tau^\pm$ spin quantization frames and the laboratory frame. As an intermediate step 
 this routine uses the
Higgs boson rest frame.
\item
   The density matrix was taken from Ref. \cite{Kramer:1994jn} and adapted to 
   the quantization 
   frames as specified in previous point. Only two cases 
    of purely scalar or pseudoscalar Higgs boson were implemented.
   Any further extension is however straightforward.
\item
    Generation of $\tau^+$ and $\tau^-$ decays  is then performed 
    following the method explained  in Ref.~ \cite{Jadach:1990mz} 
     and used in {\tt  KORALB} \cite{jadach-was:1984} since a long time ago.
\item 
   We have  assumed that  production generator provides two-body
   Higgs boson decays to $\tau$ leptons only, in particular, that it 
   does not provide any bremsstrahlung corrections. Instead,  {\tt PHOTOS} \cite{Barberio:1990ms,Barberio:1994qi}
   can be used for that purpose, once generation of $\tau^\pm$  decays is completed.
\item
   More complete inclusion of  bremsstrahlung corrections would require a 
   substantial re-write and extension of the program to the solution 
   as in Ref. \cite{kkcpc:1999} or a similar one.
\end{enumerate}

Once we have explained the main principles of our calculation, 
let us turn to the discussion of numerical results. As an example we
 will take a Higgs boson of $120$ $GeV$. In the first two  plots, which will
be constructed for the quantities defined in the Higgs boson rest-frame 
we are totally 
independent of the production
mechanism. We will study the predictions for the scalar and  pseudoscalar 
cases, essentially
to provide  the test of our generator. 
Thick lines will denote predictions for the scalar Higgs boson and 
thin lines for the pseudoscalar one. 
As in Ref.~\cite{Kramer:1994jn}, 
we  take  the  $\tau^\pm \to \nu \pi^\pm$ decay mode only.

Fig.~\ref{rysunek1}  presents the distribution in the 
 angle
$
  \phi^*=\arccos ({\vec n}^{+} \cdot {\vec n}^{-})$ where  
${\vec n}^{\pm}={ {\vec p}^{~\pi^\pm} \times  \; {\vec p}^{~\tau^-} \over
|  {\vec p}^{~\pi^\pm} \times \;  {\vec p}^{~\tau^-}|}, 
$
{\it i.e.} the  non-observable acoplanarity angle.
The  distribution is indeed, as it should be \cite{Kramer:1994jn}, 
proportional to  
$\sim 1 \mp {\pi^2 \over 16} \cos \phi^*$
respectively for scalar and pseudoscalar Higgs.
In Fig.~\ref{rysunek2} we plot the  distribution of the $\pi^+ \pi^-$ 
acollinearity angle 
($\delta^*$).
The difference between the case of a scalar and a pseudoscalar Higgs is 
clearly visible, especially
for acollinearities close to $\pi$ (see Fig.~\ref{rysunek3}).

Let us now turn to the distributions defined for the semi-realistic 
case. We need thus 
to take into consideration the combined process of decay and production 
of the Higgs boson. 
As an example%
\footnote{ We have checked that in case of other production processes 
and center-of-mass system  energies, the results, presented later 
in the paper, remain similar or are 
slightly less sensitive to 
the transverse spin ( {\it i.e.} Higgs boson parity) effects.
}%
,
for the production mechanism we took the process
$e^+e^- \to Z H$; $Z \to \mu^+\mu^-(\bar{q}q) ; H \to \tau^+ \tau^-$ (only the scalar $H$  
can be produced
in this process in ${\cal SM}$), at Center-of-Mass-System energy of $350$ $GeV$ simulated
with {\tt PYTHIA 6.1} Monte Carlo program 
\cite{Pythia}, effects due to initial state bremsstrahlung were
taken into account.  As in this case production of the pseudoscalar Higgs boson is excluded,
to quantify the size of the spin effect we will compare the predictions when all 
spin effects are included (thick lines on the following plots), with the case  
when longitudinal spin correlations are included only (thin lines).
The difference between the two lines visualizes the size of the transverse spin effects.

If we could  compare predictions for scalar and pseudoscalar, the difference would be 
roughly a factor of two larger than between the cases of full spin, longitudinal spin%
\footnote{ By numerical accident, the  case when only  longitudinal spin correlations are 
included is 
equivalent to the case of non-coherent sum of scalar and pseudoscalar Higgs boson contributions
of the equal proportions. This holds of course if the same production mechanism could be applied 
for the two cases.  
}%
. 
Then however, we could not limit our discussion
to the properties of the Higgs boson decay. Many possibilities due to generally distinct,
and model dependent, 
production mechanisms for the scalar and pseudoscalar Higgs boson would make 
the picture more involved  and not suitable for our discussion.

As we can see in Fig.~\ref{rysunek5}, the $\pi^+ \pi^-$ acollinearity angle ($\delta$) distribution 
in the laboratory frame looks quite different than in the Higgs boson rest-frame, 
the two cases of different spin treatments are indistinguishable, distribution is 
not peaked at $\delta \sim \pi$ at all.
  
If information on the beam energies and energies of all other
observed particles (high $p_T$ initial state bremsstrahlung 
photons, decay products of $Z$ {\it etc.}) are taken into considerations
the Higgs rest frame  can be reconstructed. 
We may  define the ``reconstructed'' Higgs boson  momentum as the difference of sum of beam energies
and momenta of all visible particles, that is, for  example,  
decay products of the $Z$ and all radiative photons of $| \cos\theta| < 0.98$. In our study
we will mimic in a very crude way beamstrahlung  effects only, assuming a 
flat  spread over the range  of $\pm$ $5$ $GeV$ for the longitudinal component of the Higgs boson momentum with respect to the generated one%
\footnote{ 
The typical spread  for the beam energy in linear collider is of the order of few percent
\cite{Abe:2001gc} or even worse. }.
This assumption means, that the detection effect which is practically independent from 
the Higgs boson
production mechanism is included only. 
As we can see  (Figs.~\ref{rysunek6} and ~\ref{rysunek7}) in the distribution of the acollinearity 
angle ($\delta^\bullet$) defined  
in reconstructed Higgs boson rest-frame, the effects due to transverse spin effects 
are only  barely visible.

We should stress
that, in this very simple example, we have not discussed  
at all other effects potentially 
degrading  the method, such as limited statistics, backgrounds, 
uncertainties
in  reconstruction of the 
energies and directions for the particles and jets, which may lead 
to systematic 
errors comparable in size to the parity effect, remaining after
 beamstrahlung effect is 
taken into account.  Alone, this ambiguity in reconstruction of the 
Higgs boson 
four-momentum
 degraded the method of measuring the Higgs boson parity
using the decay chain  $h \to \tau^+\tau^-$, $\tau^\pm \to \pi^\pm \nu$
 in a decisive way.  We have studied several mechanisms of the Higgs boson productions,
 in all cases depletion  of 
the acollinearity distribution sensitivity to transverse spin effect was quite similar. 
We can conclude
that our results are thus independent from the production mechanism.

 Recently
some work was started on studying detector effects, see \cite{Bower} for details.
  We can nonetheless conclude that, due to the beamsstrahlung effect, there is little hope for 
the  elegant method of reference  \cite{Kramer:1994jn}
to check Higgs boson parity using its decay to $\tau$ leptons
(whatever the luminosity of future Linear Collider), unless, may be,
other, unfortunately less sensitive to spin  than $\tau^\pm \to \pi^\pm \nu $
 decay modes are  used as well. We may hope, also that  
 methods similar to the fruitful ones for  measurement of $\tau$ polarization at LEP 1, or
for the study of CP parity and known since a long time, see {\it e.g.} 
\cite{Harton:1995dj,Nelson:1995vt},
may become available for our case as well. Definitely, realistic studies, combining accelerator,
detector and theoretical effects are needed to settle the matter.

\hspace{1 mm}
\vspace{1.5 cm}
\vskip 3 mm
\centerline{\large \bf Acknowledgements}
\vskip 3 mm
The inspiring atmosphere at the ECFA-DESY workshop in Cracow, CPP Tokyo and 
2002 Chicago Linear Collider Workshop
 was essential to the performance of this work. 
Authors are specially grateful to M. Ronan and M. Peskin for discussions.
 Useful discussion with G. Bower  is also acknowledged.

\providecommand{\href}[2]{#2}\end{document}